# Deep Learning model predicts the c-Kit-11 mutational status of canine cutaneous mast cell tumors by HE stained histological slides.

C. Puget[1*], J. Ganz[2*], J. Ostermaier[2], T. Konrad[1], E. Parlak[3], C. A. Bertram[3], M.Kiupel[5], K. Breininger[4], M. Aubreville[2], R. Klopfleisch[1]

[1]Institute of Veterinary Pathology, Free University Berlin (FUB), Germany; [2]Technische Hochschule Ingolstadt, Germany; [3]Department of Pathobiology, University of Veterinary Medicine, Vienna, Austria; [4]Department Artificial Intelligence in Biomedical Engineering, Friedrich-Alexander-Universität Erlangen-Nürnberg, Germany, [5]College of Veterinary Medicine, Michigan State University (MSU), USA, *Equal contribution

## Abstract

Numerous prognostic factors are currently assessed histopathologically in biopsies of canine mast cell tumors to evaluate clinical behavior. In addition, PCR analysis of the c-Kit exon 11 mutational status is often performed to evaluate the potential success of a tyrosine kinase inhibitor therapy.

This project aimed at training deep learning models (DLMs) to identify the c-Kit-11 mutational status of MCTs solely based on morphology without additional molecular analysis. HE slides of 195 mutated and 173 non-mutated tumors were stained consecutively in two different laboratories and scanned with three different slide scanners. This resulted in six different datasets (stain-scanner variations) of whole slide images. DLMs were trained with single and mixed datasets and their performances was assessed under scanner and staining domain shifts.

The DLMs correctly classified HE slides according to their c-Kit 11 mutation status in, on average, 87% of cases for the best-suited stain-scanner variant. A relevant performance drop could be observed when the stain-scanner combination of the training and test dataset differed. Multi-variant datasets improved the average accuracy but did not reach the maximum accuracy of algorithms trained and tested on the same stain-scanner variant.

In summary, DLM-assisted morphological examination of MCTs can predict c-Kit-exon 11 mutational status of MCTs with high accuracy. However, the recognition performance is impeded by a change of scanner or staining protocol. Larger data sets with higher numbers of scans originating from different laboratories and scanners may lead to more robust DLMs to identify c-Kit mutations in HE slides.

## Introduction

Currently, basic hematoxylin and eosin stained (HE) slides are still the standard procedure for canine cutaneous mast cell tumor (ccMCT) diagnosis. After identification of a ccMCT, a tumor grading according to Patnaik and/or Kiupel is performed.[19,12,8,25] In addition to the classic morphologic diagnosis, full diagnostic work-up of ccMCTs also includes a molecular mutation analysis of the c-Kit gene.[21]

The most commonly analyzed c-Kit mutation of canine MCTs is the internal tandem duplication (ITD) in the exon 11, which results in a constitutive c-Kit activation leading to uncontrolled cell growth and more aggressive behavior.[26,23,22] Besides the general prognostic value of information on the mutational status, confirmation of a ITD mutation of the c-Kit gene is required by the European Medicine Agency (EMA) before application of the tyrosine kinase inhibitor Masivet (Masitininb) in dogs with MCT.[16] PCR-based c-Kit mutation analysis is, however, a significant cost factor. In addition, mutation analysis is often performed on formalin-fixed and paraffin-embedded (FFPE) material which is usually affected by DNA fragmentation. This leads to a decreased sensitivity of the PCR, the current gold standard method for exon 11 ITD detection, and may completely impede molecular analysis in a non-negligible fraction of cases.[4,20]

Recent studies in human oncology have shown that deep learning models (DLMs) are able to predict the genotype of tumors solely based on their microscopic morphology in HE slides. For instance, a machine-learning algorithm was trained to predict ten specific driver mutations in human non-small cell lung carcinoma[18,5] as well as mutations in human gliomas,[13] colorectal,[9] and liver cancer[17] among others obtaining classification accuracies ranging from 0.64 to 0.93.[5,17]

This study intents to evaluate the use of DLMs for the prediction of the c-Kit exon 11 genotype of canine MCT based on HE slides. For this study, we curated a dataset of digitized HE slides of canine MCTs with PCR confirmation of the c-Kit exon 11 genotype.

To reflect the diversity of HE-staining variants and scanner types in diagnostic institutions, each slide was stained in two different labs and scanned with three different scanning devices, resulting in six different data sets (stain-scanner-variants). DLMs were either trained and evaluated on the same dataset they were trained on (same stain-scanner variant, in-domain testing) or trained on a single/mixed dataset and tested on an unknown dataset (different stain-scanner-variant, out-of-domain).

**Material and Methods**

*Tumors*

The study sample was composed of 368 HE stained histological slides of MCT biopsies that were sent to the Veterinary Laboratory of the Michigan State University (MSU) between 2018 and 2022 for morphological grading as well as molecular evaluation of the mutational status of the c-Kit exon 11. The histological grade (according to Patnaik[19] and/or Kiupel[12]) was gathered, if available, from of the histopathological reports, or attributed retrospectively by 3 board-certified veterinary pathologists. Further information such as breed, age, gender, localization of the MCT, involvement of the different skin layers (dermal, subcutaneous or mucocutaneous) and surface condition (ulcerated or intact) could be collected in most of the cases (see Supplemental Table 1).

Thick tissue sections (several cell layers in fine focus), tumors with extensive necrotic tumor area, specimens in bad conservation state, damaged slide glass, presence of artefacts under the coverslip (air bubbles, tissue folds), and missing information about the c-Kit mutational status of exon 11 were exclusion criteria.

In total, 50 different dog breeds were represented in the data set. The 10 most occurring breeds, accounting for 198 cases (53.8%), were: Mixed breed (37.4%), Labrador Retriever (23.7%), Golden Retriever (8.6%), Boxer (6.1%), American Pit Bull Terrier (5.6%), Maltese Dog (4.5%), Pug (4.5%), French Bulldog (3.5%), Boston Terrier (3.5%) and Shih Tzu (2.5%). (Supplemental Table 2). Breed information was unavailable for 96 (26.1%) cases.

Out of the selected 195 c-Kit-11 mutated MCTs, 121 (62.1%) were Kiupel high grade and 74 (37.9%) were Kiupel low grade. 164 (84.1%) were cutaneous, 18 (9.2%) were subcutaneous, 6 (3.1%) were mucocutaneous and 7 (3.6%) could not be reliably located due to lacking histological report and absence of orientation criteria on the slide. The 173 c-Kit non-mutated MCTs were chosen in the aim of mirroring the repartition of the c-Kit-11 mutated group, with 97 (56.1%) being classified as high grade, and 76 (43.9%) low grade. Moreover, 145 (83.8%) were cutaneous, 26 (15.0%) subcutaneous, and 2 (1.2%) were mucocutaneous (see Table 1).

**Table 1:** Mast cell tumors, mutational status, Kiupel and Patnaik Grade

|  | c-Kit mutated | | | | c-Kit wild type | | | |
|---|---|---|---|---|---|---|---|---|
| n | 195 | | | | 173 | | | |
| Kiupel grade | High | | Low | | High | | Low | |
| n | 121 | | 74 | | 97 | | 76 | |
| Patnaik grade | 3 | 2 | 2 | 1 | 3 | 2 | 2 | 1 |
| n | 55 | 66 | 5 | 69 | 35 | 62 | 70 | 6 |

*PCR-based identification of c-Kit exon 11 Mutational status*

Neoplastic cells were isolated from FFPE MCTs with laser-capture microdissection (LCM) as previously described.[24] The exon 11 of the c-Kit gene was amplified by PCR using the primer pair at the 5' end of the exon 11 (PE1: 5'-CCATGTATGAAGTACAGTGGAAG-3' sense, bp 1,657–1,680 of exon 11) and the 5'-end of intron 11 (PE2: 5'GTTCCCTAAAGTCATTGTTACACG-3' anti-sense, nucleotides 43–66 of intron 11).[3] A Mutation Detection PCR Kit (Qiagen) was used for the amplification.[2] PCRs were prepared in a 25-µL total reaction volume, with 10 to 25 ng of LCM-extracted DNA and 5 pmol of each primer 0.5 U of Taq polymerase (Invitrogen, Carlsbad, CA), and final concentrations of 80 µM deoxynucleoside triphosphate, 2 mM MgCl2, 20 mM Tris–HCl, and 50 µl of KCl.[24] Cycling conditions were as follows: 95 °C for 5 minutes; 35 cycles at 95 °C for 30 seconds, 58 °C for 90 seconds, 72 °C for 30 seconds; followed by 68 °C for 10 minutes.[2] PCR products were visualized on the QIAxcel Capillary Electrophoresis System (Qiagen).[2]

*Staining variants*

To investigate the dependency of the mutational status prediction on the staining protocol, we applied two stains (Stain A, Stain B) to the same tissue section, which was de-stained in between both staining steps, using the following protocol:

**HE-stain A** – Paraffin sections of all tumors were originally stained at MSU Veterinary Diagnostic Laboratory using the VENTANA HE600 system from Roche. Dry deparaffinization, rehydration and HE staining are run according to standard protocol[7] and using VENTANA HE 600 Hematoxylin (Roche, 07024282001) and VENTANA HE 600 Eosin (Roche, 06544304001).

**De-staining** – After scanning with all three scanners (see below), all slides were manually de-stained. First, the glass slides were immersed in xylol for at least 2 days before carefully manually removing the coverslip. They were then transferred into a 1% hydrochloric acid alcohol solution until all color was washed away and subsequently rinsed with distilled water.

*HE-stain B* – De-stained slides were then manually dipped in Mayer's Hemalaun (Carl Roth, T865.3) for 8 minutes, rinsed in tap water, dipped in eosin (Diagonal, 2C140.01000) for 30 seconds and shortly rinsed in tap water before going through an ascending alcohol series (1 minute in 70%, 80%, 96% and 100% alcohol, respectively).

*Scanners and stain-scanner variant datasets*

All 368 slides were scanned twice (with stain A and consecutively with stain B) on three different scanners, Aperio CS2 (scanner 1), Aperio AT2 scanner (scanner 2) and 3DHistech Pannoramic II scanner (scanner 3). This resulted in in six distinct stain-scanner variant datasets of WSI (see Figure 1): dataset 1 (stain A + scanner 1), dataset 2 (stain A + scanner 2), dataset 3 (stain A + scanner 3), dataset 4 (stain B + scanner 1), dataset 5 (stain B + scanner 2), dataset 6 (stain B + scanner 3).

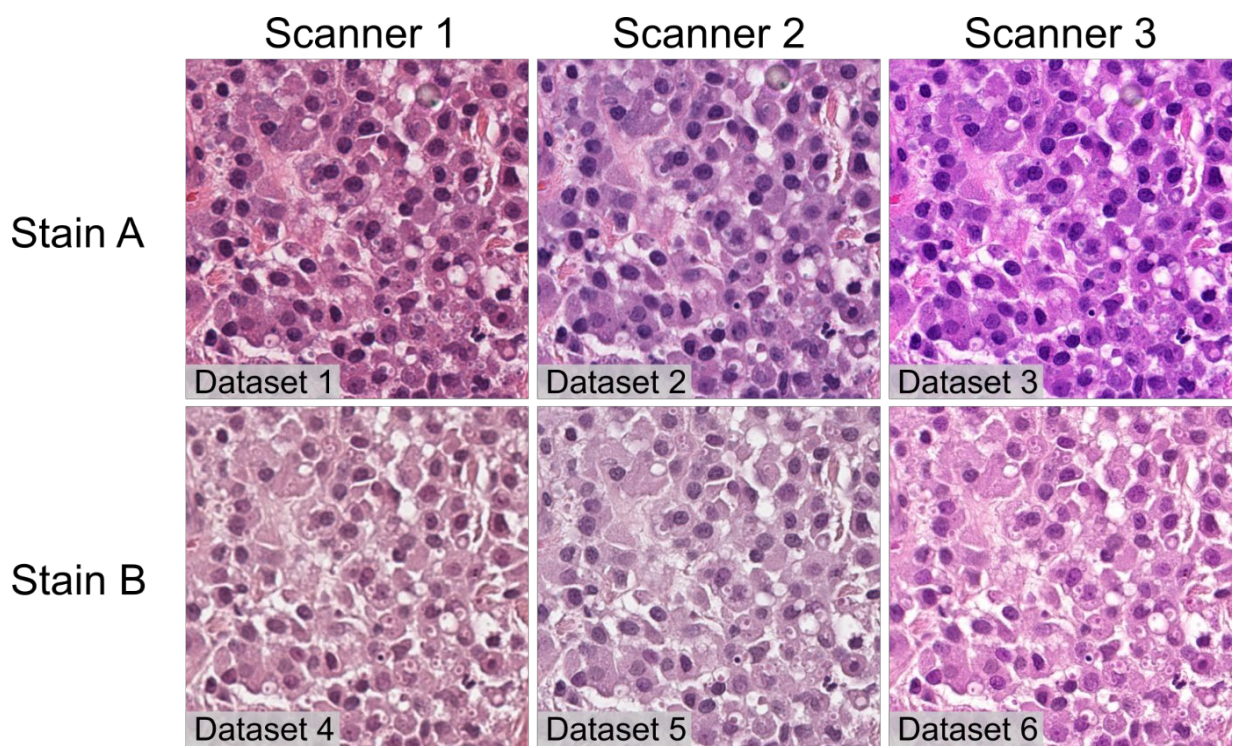

**Figure 1.** Impact of stain and scanner type on WSI characteristics. Images of the same tumor sample with six stain-scanner variants, mast cell tumor, skin, canine, HE. Each image represents the same portion of the same tumor in a different stain-scanner variant. Images produced by the different scanners (comparison in the same row) show differences in color calibration and a slight difference in sharpness / depth of field. The stainings differ in colors shades / intensity and in the amount of visible detail (comparison in the same column).
Scanner 1 = Aperio CS2; Scanner 2 = Aperio AT2; Scanner 3 = 3DHistech Pannoramic II.

*Deep learning model*

Since WSIs are typically too large to fit into GPU-memory during training, smaller image patches are extracted from the WSI, and the patch-level information is aggregated to form a slide-level prediction (Fig. 2).

In this work, this was done by formulating the problem as a multiple-instance learning (MIL) task. In MIL, the WSI is split into image patches (instances), which together are commonly referred to as a *bag of instances*.[10] The class label is only provided on the slide (bag) level, as the c-Kit mutational status is not known for each individual patch. The Clustering-Constrained Attention Multiple Instance Learning (CLAM) algorithm by Lu et al.[15] was used for multiple instance learning. In this framework, each patch is assigned an attention score (AS) that directly expresses the importance of the patch to the model's prediction. The information contained in the patches is weighted according to the attention score and aggregated over the slide. This way, the complete slide (bag) is used for predicting the label, i.e., the mutational status.

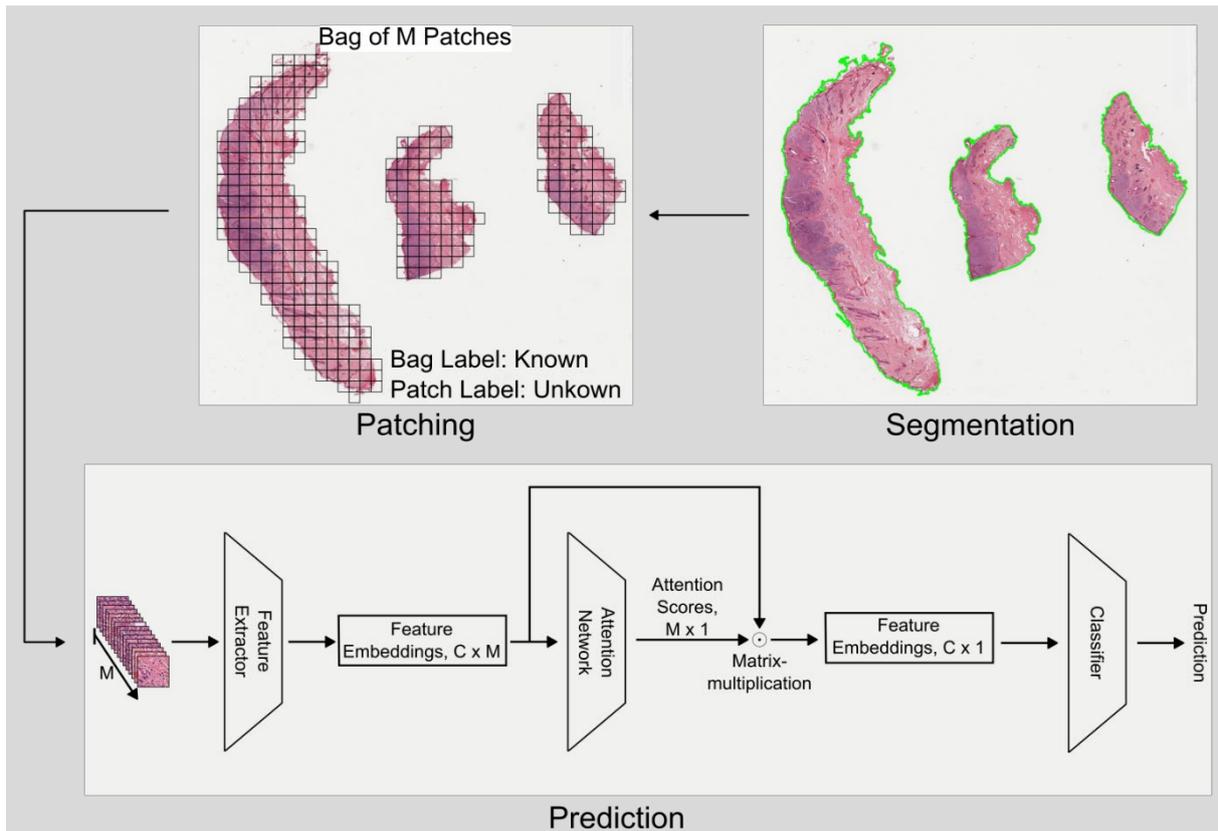

**Figure 2**. Schematic representation of the Multiple Instance Learning (MIL) and Clustering-Constrained Attention Multiple Instance Learning (CLAM) concept.
WSI classification is performed in several steps within the CLAM pipeline used in this study.
First, the tissue is segmented and then tissue region is separated into M patches. For each patch, a feature embedding of size C x 1 is computed, resulting in a vector of size C x M. Prediction of this information (if a patch of the image contributes to the discrimination or not) is the target of the attention score calculation. Patches that do not contribute shall hence be assigned an attention value of zero and will be discarded for the decision-making. This makes the attention score directly interpretable. The aggregation to a single feature embedding of size C x 1 is then performed by matrix multiplication. This feature embedding is representative for the whole slide and is used for the final prediction of the mutational status.

*Model Training Scheme*

To increase the statistical informativeness of the results, the DLMs were trained using a 10-fold Monte Carlo cross-validation. This means, that for each of ten consecutive runs, the dataset was randomly split into disjoint training-, validation, and test splits. Within each split, 85 % of the slides were used as training and validation data, and 15% were used as test data. Afterward, in each run a model was trained on the training data. To ensure comparability across experiments, the same ten trainings, validations, and test splits were used for the experiments across all dataset variants. All DLMs were trained until convergence as observed by the validation loss. Implementation details of the used model architecture and training parameters can be found in Appendix A-1.

*Training approach – Single dataset*

In a first approach, DLMs were trained using the training fraction of the six datasets. Each DLM was then tested on the test fraction of all datasets, resulting in one in-domain and five out-of-domain tests (Fig.3, left side). Subsequently we will refer to this approach as **single dataset training**.

*Training approach - Mixed datasets ("Leave-one-out approach")*

To assess whether the training with a more diverse dataset leads to a more robust prediction of the c-Kit 11 mutational status, DLMs were trained with combined training datasets composed of five of the six original datasets, i.e. leaving one dataset out of the training. The DLMs were then tested on the test fraction of all datasets, resulting in five in-domain and one out-of-domain tests. (Fig.3, right side). In the following paragraphs we will refer to this approach as **mixed dataset training**.

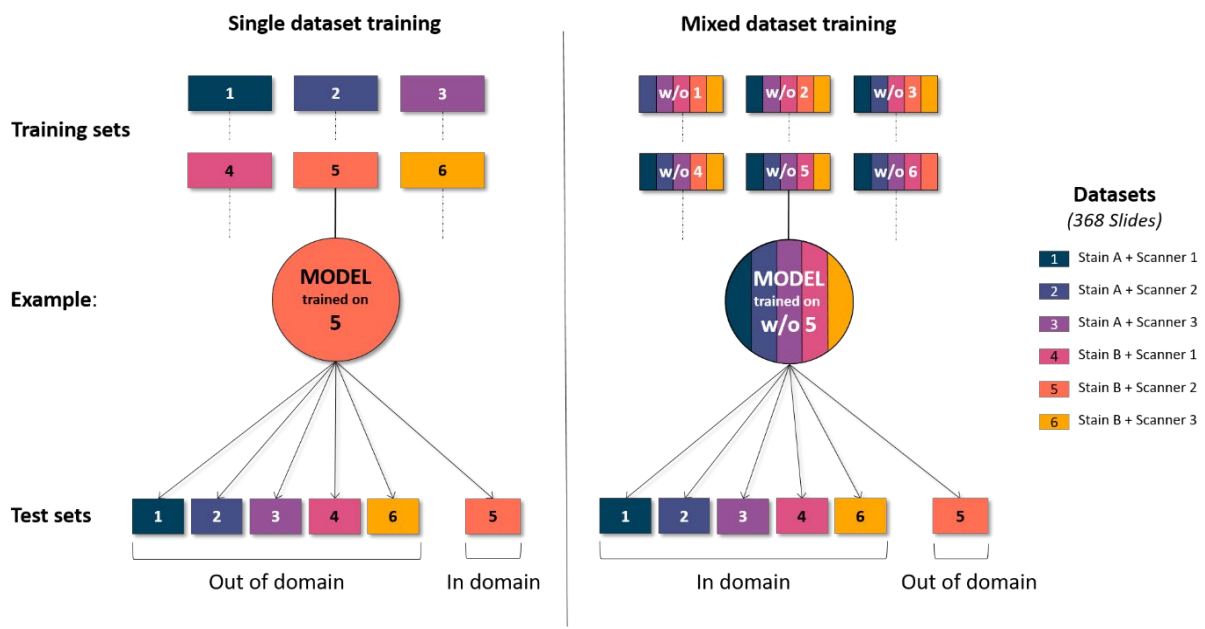

**Figure 3.** Schematic representation of the c-Kit-11 detection model training approaches. Two general training strategies were applied for DLM training. Training sets composed of single datasets (left side) and training sets composed of equal shares of five of the six datasets (right side) were used. The resulting trained DLMs were consecutively tested on all six datasets to assess their performances in-domain

(tested on a dataset from the same stain-scanner variant) and out-of-domain (tested on a stain-scanner variant that was not included in the training).
w/o = without; Scanner 1 = Aperio CS2; Scanner 2 = Aperio AT2; Scanner 3 = 3DHistech Pannoramic II.

*Test for independence of diagnostic variables*

To assess the association between the model's predictions and clinical-pathological variables, like the Patnaik or Kiupel tumor grade, we used the Chi-square test of independence. This is a statistical method for assessing the associations between variables of different scales. To lower the chances of misinterpreting two variables as associated, we applied Bonferroni correction since we conducted multiple statistical tests simultaneously. The assessment only considered models trained on a single dataset to establish a direct link between training data and DLM results. When a mixed dataset is used for training, it is more difficult to link the association between DLM predictions and clinical-pathological variables to a specific dataset that is part of the mixed dataset.

**Results**

*Prediction of the c-kit 11 mutational status using single dataset training*

To assess the capacity of an DLM trained in a single stain-scanner environment to predict the c-Kit-11 mutational status in HE-stained WSIs, six DLMs were trained on a single training set (stain-scanner-variant) and tested on a training set of the same stain-scanner variant (in-domain) or a different stain-scanner variant (out-of-domain) (Table 2).

Across all DLMs an in-domain average mean classification accuracy (MCA) of 0.79 (range 0.75-0.87) was found. The highest MCA of 0.87 was reached by the DLM trained on the dataset 6 (stain B, scanner 3) with a sensitivity of 0.90 and a specificity of 0.83. In general, DLMs trained and tested on stain B WSIs allowed for a higher MCAs (average MCA 0.82) than DLMs trained on stain A WSIs (average MCA 0.76), pointing towards a general influence of HE-staining protocol for accurate c-Kit-11 mutation prediction. Furthermore, scanner 3 seemed to be slightly more suitable for c-Kit-11 mutation prediction (average MCA 0.84) than scanner 1 (average MCA 0.78) and scanner 2 (average MCA 0.77).

However, when the models were tested out-of-domain, they only reached an average MCA of 0.65 (range 0.54-0.76).

**Table 2.** Mean classification accuracies and standard deviations of c-Kit 11 prediction with DLMs trained on single datasets.

| Tested on dataset | Trained on dataset | | | | | |
|---|---|---|---|---|---|---|
| | 1 | 2 | 3 | 4 | 5 | 6 |
| 1 | **0.75** ± 0.08 | 0.63 ± 0.04 | 0.68 ± 0.08 | 0.59 ± 0.04 | 0.61 ± 0.04 | 0.62 ± 0.07 |
| 2 | 0.74 ± 0.05 | **0.75** ± 0.07 | 0.68 ± 0.05 | 0.71 ± 0.05 | 0.70 ± 0.04 | 0.68 ± 0.04 |
| 3 | 0.73 ± 0.06 | 0.66 ± 0.03 | **0.80** ± 0.04 | 0.72 ± 0.06 | 0.57 ± 0.04 | 0.58 ± 0.05 |
| 4 | 0.63 ± 0.08 | 0.61 ± 0.06 | 0.61 ± 0.06 | **0.81** ± 0.05 | 0.76 ± 0.06 | 0.61 ± 0.08 |
| 5 | 0.63 ± 0.06 | 0.72 ± 0.07 | 0.61 ± 0.06 | 0.74 ± 0.05 | **0.78** ± 0.05 | 0.67 ± 0.09 |
| 6 | 0.65 ± 0.07 | 0.58 ± 0.04 | 0.54 ± 0.08 | 0.75 ± 0.05 | 0.59 ± 0.03 | **0.87** ± 0.05 |

Bold font = mean classification accuracy of DLMs trained and tested on the same dataset (in-domain); regular font = mean classification accuracy of DLMs tested on a dataset different training dataset (out-of-domain).

*Prediction of the c-kit 11 mutational status using mixed dataset training (leave one out approach)*

To test whether more diverse datasets from multiple institutions may lead to more robust c-Kit-11 prediction in HE-stained WSI from unknown stain-scanner variants, a second set of six DLMs was trained on a dataset that contained equal fractions (20%) of five of the six datasets and tested on the sixth, unknown, dataset (leave-on-out-approach, Table 3).

The average in-domain MCA was 0.76 (range 0.72-0.85), which is lower than the in-domain MCA of the DLMs trained and tested on a single dataset (Table 2).

However, with an average MCA of 0.73 (range 0.69-0.77), the models trained on mixed datasets showed a better out-of-domain performance than the models trained on a single dataset, which achieved an average out-of-domain MCA of 0.65 (range 0.54-0.76) (Table 2).

**Table 3.** Mean classification accuracies and standard deviations of c-Kit 11 prediction with DLM trained on mixed datasets (leave-one-out-approach).

| Tested on dataset | Trained on dataset | | | | | |
|---|---|---|---|---|---|---|
| | w/o 1 | w/o 2 | w/o 3 | w/o 4 | w/o 5 | w/o 6 |
| 1 | **0.69** ± 0.04 | 0.74 ± 0.05 | 0.73 ± 0.06 | 0.73 ± 0.07 | 0.72 ± 0.06 | 0.73 ± 0.07 |
| 2 | 0.76 ± 0.06 | **0.75** ± 0.06 | 0.76 ± 0.06 | 0.76 ± 0.07 | 0.75 ± 0.07 | 0.75 ± 0.08 |
| 3 | 0.77 ± 0.06 | 0.76 ± 0.05 | **0.71** ± 0.05 | 0.77 ± 0.06 | 0.76 ± 0.04 | 0.79 ± 0.06 |
| 4 | 0.80 ± 0.05 | 0.77 ± 0.05 | 0.75 ± 0.05 | **0.72** ± 0.05 | 0.75 ± 0.06 | 0.76 ± 0.06 |
| 5 | 0.79 ± 0.06 | 0.77 ± 0.05 | 0.77 ± 0.05 | 0.77 ± 0.06 | **0.77** ± 0.05 | 0.77 ± 0.06 |
| 6 | 0.85 ± 0.04 | 0.82 ± 0.04 | 0.83 ± 0.04 | 0.82 ± 0.05 | 0.83 ± 0.04 | **0.73** ± 0.05 |

Bold font = mean classification accuracy (MCA) of DLMs tested on a dataset different from the one they were trained on (out-of-domain); regular font = classification accuracy of DLMs trained and tested on the same dataset (in-domain); w/o = without.

*Multivariate analysis of c-Kit mutation prediction, c-Kit mutation status and other clinic-pathological parameters*

A multivariate analysis was used to test for interdependence between the DLM-based mutation prediction and the following parameters: location of the tumor (subcutaneous, cutaneous, mucocutaneous), skin ulceration, tumor grade (Kiupel and Patnaik grading system) and sex (Table 4 and Supplemental table 3).

Analysis confirmed that the DLM predictions were not significantly interdependent with either the location or the sex of the affected dog. A significant association was found between the Patnaik grade and the predictive output of all models trained on stain A (DLMs trained on datasets 1-3) as well as between the c-Kit 11-mutation prediction by the DLM trained on dataset 1 and 4 and the Kiupel grade. Chi-square analysis indicated that the ulceration status of the overlying epidermis of the skin is significantly associated with the mutation prediction by all DLMs (trained on datasets 1-6).

**Table 4.** Associations between the DLM prediction and known clinical-pathological parameters of the tumor.

| Parameter | PCR | DLM trained on dataset | | | | | |
|---|---|---|---|---|---|---|---|
| | | 1 | 2 | 3 | 4 | 5 | 6 |
| Location | - | - | - | - | - | - | - |
| Ulcerated surface | + | + | + | + | + | + | + |
| Kiupel grade | + | + | - | - | + | - | - |
| Patnaik grade | - | + | + | + | - | - | - |
| Sex | - | - | - | - | - | - | - |

P-values < 0.05 are marked with "+" and p-values > 0.05 with "-".

**Discussion**

The study aimed at testing whether Deep Learning Models (DLMs) are able to predict the c-Kit exon 11 mutational status of canine cutaneous mast cell tumors (MCTs) based exclusively on HE-stained WSI. To reflect the variability of staining protocols and scanning devices used in different institutions, the slides were consecutively stained with two HE stains from different laboratories and digitalized with three different slide scanners. This resulted in six datasets of HE-stained Whole Slide Images (WSIs).

DLMs were either trained on one of the six different scanner-stain combinations (single dataset), or on mixtures of scanner-stain combinations (mixed dataset). Training with a single dataset corresponds to the situation in which an institution trains a DLM based on its own archive data. On the contrary, training with a mixed dataset corresponds to several institutions contributing to the training of a DLM with combined archive data.

Evaluation of the cross-validation results of the single dataset trained DLMs tested on the test portion of the same datasets they were trained on (in domain, i.e., from the same diagnostic laboratory) showed that the DLMs were able to predict the c-Kit-11 mutation status with a mean classification accuracy (MCA) of up to 0.87 (dataset 6, sensitivity of 0.90 and a specificity of 0.83). Across the single dataset trained DLMs, the average MCA was 0.79. Prediction of c-Kit-11 exon 11 mutation is thus possible in general with high accuracy, given a suitable staining protocol and scanner device.

Further analysis showed, that stain B allowed a better c-Kit mutation prediction with all three scanners. This stain therefore seems to highlight the, so far unknown, morphologic features associated with c-Kit-11 mutation better than stain A.

Testing of the single dataset trained DLMs on WSIs with a different stain-scanner variant (out-of-domain, i.e. from a different diagnostic laboratory) led to a relevant MCA drop, yielding an average MCA of 0.65 and a maximum MCA of 0.76. Transfer of a DLM trained in one laboratory to another laboratory may thus be limited by the choice of scanner hardware and staining protocol. This high domain-dependency of DLMs has been described for HE-stain WSI before[1] and domain generalization is currently a major research focus in computer science to reduce data set variability required for robust AI applications.

The chi-square analysis suggests varying associations between DLM-based predictions and clinical-pathological variables based on different training datasets. This implies that prediction of the c-Kit-11 mutational status is based on morphological features that are specific to a single stain-scanner combination. Hence, a model trained on one stain-scanner combination may not generalize to other combinations without specific adjustments. In general, the morphologic features in WSIs, which allow DLMs to predict c-Kit 11 mutation, were not identified in the present study. However, it would be of interest to determine which structures allow for c-Kit-11 mutation detection in HE slides and whether these structures can be applied by human observers. Identifying these structures are subject to current studies of our group.

In a second training approach, DLMs were trained on datasets composed of varying assortments of five of the six original datasets (mixed dataset). These were tested on the five known in-domain and the sixth out-of-domain datasets to simulate the transfer of these mixed training DLM into an unknown laboratory. Tested in-domain, these DLMs predicted the c-Kit-11 mutational status with an only mildly lower average MCA of 0.76 compared to the single scanner-stain variation-trained DLMs. However, this approach led on average to a mildly higher out of domain MCA of 0.73 than the single dataset trained DLMs tested out of domain.

In conclusion, DLMs can predict the c-Kit-11 mutational status of canine cutaneous MCTs based on HE-stained WSIs with an MCA of up to 0.87. Further studies are necessary to increase the MCA and train more field ready DLMs in the prospect of saving the time and related costs of the PCR based diagnostic. The capacity of human observers to predict the c-Kit-11 mutational status after a DLM-assisted training remains to be assessed.

**Author Note**

Chloé Puget and Jonathan Ganz contributed equally to the study.

**Declaration of Conflicting Interests**

The authors declared no potential conflicts of interest with respect to the research, authorship, and/or publication of this article.

**Funding**

The authors received no financial support for the research, authorship, and/or publication of this article.

**ORCID iD**

Chloé Puget: https://orcid.org/0000-0001-7629-0356

Jonathan Ganz: https://orcid.org/0009-0008-1299-8716

Christof A. Bertram: https://orcid.org/0000-0002-2402-9997

Matti Kiupel: https://orcid.org/0000-0003-4429-9529

Katharina Breininger: https://orcid.org/0000-0001-7600-5869

Marc Aubreville: https://orcid.org/0000-0002-5294-5247

Robert Klopfleisch: https://orcid.org/0000-0002-6308-0568

# Appendix

## A-1 Implementation and Training Details

In contrast to the original CLAM pipeline, which uses a 50-layer residual network (ResNet50) for feature extraction, we used the state-of-the-art ConvNext[14] model pre-trained on the ImageNet database[6] for feature extraction. This model is reported to outperform the ResNet50 in several downstream tasks when used as a feature extractor. The network details are as follows: The feature dimension was reduced using a projection layer before the attention network. This layer comprised of one linear layer with 1024 input and 512 output nodes, followed by a rectified linear unit (ReLU) activation function. The attention network consisted of two linear layers with a hyperbolic tangent activation function in-between. The first layer had 512 input and 256 output nodes, and the second layer had 256 input and one output node. The final classification network included a single linear layer with 512 input nodes and 2 output nodes. For computational efficiency, all models were trained with the Adam optimizer[11] for a minimum of 60 epochs and a maximum of 200 epochs at a learning rate of 0.0001. The best model from each training session was retrospectively selected based on its validation loss. All trainings were conducted on a single NVIDIA RTX 3090 GPU.